\documentclass[twocolumn]{jpsj3}

\usepackage{color}
\usepackage{bm}

\title{%
Chiral Surface States on the Step Edge in a Weyl Semimetal
}

\author{%
Yositake Takane
}

\inst{%
Department of Quantum Matter, Graduate School of Advanced Sciences of Matter,\\
Hiroshima University, Higashihiroshima, Hiroshima 739-8530, Japan
}

\recdate{ \hspace{50mm} }

\abst{%
A Weyl semimetal with a pair of Weyl nodes accommodates chiral states
on its flat surface if the Weyl nodes are projected onto two different points
in the corresponding surface Brillouin zone.
These surface states are collectively referred to as a Fermi arc
as they appear to connect the projected Weyl nodes.
This statement assumes that translational symmetry is present
on the surface and hence electron momentum is a conserved quantity.
It is unclear how chiral surface states are modified
if the translational symmetry is broken by a particular system structure.
Here, focusing on a straight step edge of finite width,
we numerically analyze how chiral surface states appear on it.
It is shown that the chiral surface states are algebraically (i.e., weakly)
localized near the step edge.
It is also shown that the appearance of chiral surface states
is approximately determined by a simple condition characterized by
the number of unit atomic layers constituting the step edge
together with the location of the Weyl nodes.
}

\begin{document}
\sloppy
\maketitle

\section{Introduction}

A Weyl semimetal possesses pairs of nondegenerate Dirac cones
with opposite chirality.~\cite{volovik,murakami,
wan,yang,burkov1,burkov2,WK,delplace,halasz,sekine}
The band-touching point of each Dirac cone is referred to as a Weyl node.
A typical feature of a Weyl semimetal is that low-energy states
with chirality appear on its flat surface~\cite{wan}
if a pair of Weyl nodes is projected onto two different points
in the corresponding surface Brillouin zone.
As these surface states appear to connect a pair of
projected Weyl nodes, they are collectively called a Fermi arc.
The presence of a Fermi arc gives rise to
an anomalous Hall effect.~\cite{burkov1}
It has been shown experimentally that a Weyl semimetal phase is realized in
TaAs and NbAs.~\cite{weng,huang1,xu1,lv1,lv2,xu2}

Usually, the appearance of chiral surface states is argued
under the assumption that translational symmetry is present on the surface
and hence electron momentum is a conserved quantity.
How are chiral surface states modified if the translational symmetry is
explicitly broken as a consequence of a particular system structure?
Let us focus on a straight step arranged on the flat surface
of a Weyl semimetal [see Fig.~1(a)].
On its side surface, which is referred to as a step edge hereafter,
the translational symmetry is broken in the perpendicular direction.
It has been shown that in a weak topological insulator, surface states
on such a step edge exhibit unusual properties.~\cite{yoshimura1,arita1,pauly1,
matsumoto,zhou,arita2,pauly2}
Let $N$ be the number of unit atomic layers constituting the step edge.
If $N$ is very large, the translational symmetry is approximately recovered
on the step edge and hence surface states behave in the same way as
on an infinitely large flat surface.
We thus concentrate on the case where $N$ is relatively small,
typically of order ten or smaller.
\begin{figure}[btp]
\begin{center}
\includegraphics[height=2.5cm]{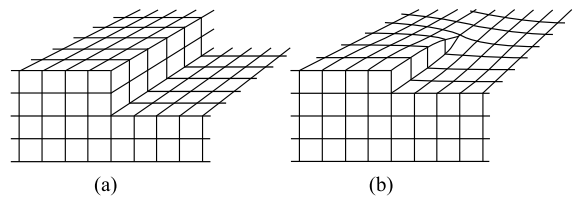}
\end{center}
\caption{
(a) Straight step edge of $N = 2$. (b) Step edge of $N = 1$
induced by a screw dislocation.
}
\end{figure}

For definiteness, we focus on the simple system of a Weyl semimetal
on a cubic lattice with a pair of Weyl nodes
at $\mib{k}_{\pm} = (0,0,\pm k_{0})$.
On a flat surface parallel to the $xy$-plane,
chiral surface states disappear as $\mib{k}_{\pm}$ are projected onto
the same point [i.e., $(k_{x},k_{y}) = (0,0)$]
in the corresponding surface Brillouin zone.
Contrastingly, chiral surface states appear on a surface parallel
to the $yz$- or $zx$-plane.
Let us assume that a straight step consisting of $N$ unit atomic layers
is arranged on the flat surface parallel to the $xy$-plane
so that a step edge of height $N$ appears
with its surface perpendicular to the $xy$-plane.
Note that this step edge can be related to a screw dislocation
along the $z$-direction with a displacement of $N$ unit atomic layers.
If such a screw dislocation is terminated on a flat surface, a step edge of
height $N$ originating from the point of termination on the surface
inevitably appears [see Fig.~1(b)].
It has been shown that chiral states appear along a screw dislocation
in a Weyl semimetal under a certain condition.~\cite{imura1}
We can expect that if a chiral state is stabilized along the dislocation,
it is continuously connected with a chiral surface state on a step edge.
Adapting the argument of Ref.~\citen{imura1} to this case, we find that
a pair of gapless chiral states is stabilized along a screw dislocation
for each odd integer $l$ satisfying $k_{0}a > l\pi/N$.
Do chiral surface states on a step edge satisfy the same condition?
This condition is different from that for chiral surface states
in the slab of a Weyl semimetal.
If the slab consists of $N$ unit atomic layers stacked in the $z$-direction,
an unpaired chiral surface state appears on its side surface
for each positive integer $l$ satisfying
$k_{0}a > l\pi/(N+1)$.~\cite{takane,yoshimura2}

In this paper, we study the behavior of chiral surface states
on a straight step edge of height $N$ in a prototypical Weyl semimetal
possessing a pair of Weyl nodes at $\mib{k}_{\pm} = (0,0,\pm k_{0})$.
We numerically obtain the wave functions of low-energy eigenstates
near the Weyl nodes in the model system with two step edges.
It is shown that the chiral surface states are algebraically (i.e., weakly)
localized near the step edge.
This should be contrasted with the strongly localized feature of
chiral states on a flat surface or along a screw dislocation.
It is also shown that the appearance of chiral surface states is approximately
described by a simple condition: an unpaired chiral surface state appears
on the step edge for each positive integer $j$ satisfying $k_{0}a > j\pi/N$.
This condition is different from
that for chiral states along a screw dislocation.

In the next section, we present a tight-binding model for a Weyl semimetal
and implement it on a lattice system with two step edges.
In Sect.~3, we numerically obtain
the wave functions of low-energy states near the Weyl nodes.
By analyzing the resulting wave functions, we clarify the appearance
of chiral surface states on a step edge.
The last section is devoted to a summary.

\section{Model and Numerical Method}

For a Weyl semimetal with a pair of Weyl nodes
at $\mib{k}_{\pm} = (0,0,\pm k_{0})$, we introduce a tight-binding model
on a cubic lattice with the lattice constant $a$.
The energy at the Weyl nodes is set equal to zero.
As our attention is restricted to the system being infinitely long
in the $x$-direction, we hereafter characterize the $x$ dependence of
each eigenstate with a wave number $k_{x}$.
The indices $m$ and $n$ are respectively used to specify lattice sites
in the $y$- and $z$-directions,
and the two-component state vector for the $(m,n)$th site is expressed as
\begin{align}
  |m,n \rangle
  =  \left[ |m,n \rangle_{\uparrow}, |m,n \rangle_{\downarrow}
     \right] ,
\end{align}
where $\uparrow, \downarrow$ represents the spin degree of freedom.
The tight-binding Hamiltonian is given by
$H = H_{0}+H_{y}+H_{z}$ with~\cite{yang,burkov1}
\begin{align}
   H_{0}
 & = \sum_{m,n} |m,n \rangle h_{0} \langle m,n| ,
         \\
   H_{y}
 & = \sum_{m,n}
     \left\{ |m+1,n \rangle h_{y}^{+} \langle m,n| + {\rm h.c.} \right\} ,
         \\
   H_{z}
 & = \sum_{m,n}
     \left\{ |m,n+1 \rangle h_{z}^{+} \langle m,n| + {\rm h.c.} \right\} .
\end{align}
Here, the $2 \times 2$ matrices are
\begin{align}
   h_{0}
 & = \left[ 
       \begin{array}{cc}
         \zeta(k_{x}) & A\sin(k_{x}a) \\
         A\sin(k_{x}a) & -\zeta(k_{x})
       \end{array}
     \right] ,
               \\
   h_{y}^{+}
 & = \left[ 
       \begin{array}{cc}
         -B & \frac{1}{2}A \\
         -\frac{1}{2}A & B
       \end{array}
     \right] ,
               \\
   h_{z}^{+}
 & = \left[ 
       \begin{array}{cc}
         -t & 0 \\
         0 & t
       \end{array}
     \right] ,
\end{align}
with
\begin{align}
  \zeta(k_{x}) = 2t\cos(k_{0}a) + 2B(2-\cos(k_{x}a)) ,
\end{align}
where $\pi > k_{0}a > 0$, and the other parameters, $A$, $B$, and $t$,
are assumed to be real and positive.
We find that the energy dispersion of this model is
\begin{align}
           \label{eq:exp-E}
   E =
 & \pm \Big\{\left[\Delta(k_{z})
                   +2B\bigl(2-\cos(k_{x}a)-\cos(k_{y}a)\bigr)\right]^{2}
              \nonumber \\
 & \hspace{10mm}
           + A^{2}\bigl(\sin^{2}(k_{x}a)+\sin^{2}(k_{y}a)\bigr)
       \Big\}^{\frac{1}{2}}
\end{align}
with
\begin{align}
  \Delta(k_{z}) = - 2t\left[\cos(k_{z}a)-\cos(k_{0}a)\right] .
\end{align}
Equation~(\ref{eq:exp-E}) indicates that a pair of Weyl nodes
appears at $\mib{k}_{\pm} = (0,0,\pm k_{0})$, where $E = 0$.

\begin{figure}[btp]
\begin{center}
\includegraphics[height=4.5cm]{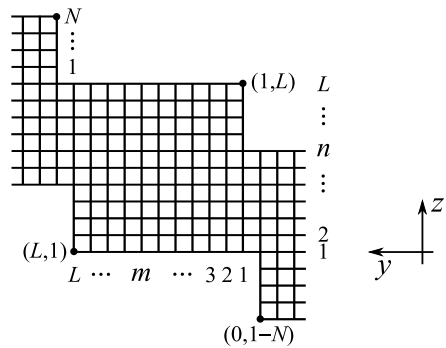}
\end{center}
\caption{
Cross section of the system considered in the text;
it is constructed from a set of $L \times L$ sites
by imposing a shifted periodic boundary condition in the $y$-direction
with a shift of $N$ sites.
The system is infinitely long in the $x$-direction and possesses
two straight step edges of height $N$: the lower left step edge includes
the corner site at $(L,1)$ while the upper right one includes that at $(1,L)$.
Note that the corner site at $(0,1-N)$ is identical with
that at $(L,1)$ under the shifted periodic boundary condition.
}
\end{figure}
Let us introduce the lattice system with the cross section (see Fig.~2)
being constructed from the set of $L \times L$ sites
under a shifted periodic boundary condition in the $y$-direction
with the shift of $N$ sites.
This is infinitely long in the $x$-direction
and possesses two straight step edges of height $N$.
We apply the tight-binding model given above to this lattice system.
As the top and bottom surfaces are parallel to the $xy$-plane,
chiral surface states appear only on the lower left and upper right step edges
parallel to the $zx$-plane.
The direction of propagation on one step edge
is opposite to that on the other step edge.

As chiral surface states should appear near the Weyl nodes at $E = 0$,
we thus numerically obtain the wave functions of eigenstates
near zero energy (i.e., low-energy states).
We find that the lowest-energy state, as well as lower-energy states
in some cases, is algebraically localized near the step edge.
That is, a probable chiral surface state is only weakly localized
near the step edge.
In this situation, we need to examine
whether such an algebraically localized state can be a true surface state.
Let us suppose that an eigenstate is localized near one step edge
when the system size $L$ is large but finite.
If its wave function is exponentially (i.e., strongly) localized
near the step edge, it is insensitive to the variation of $L$.
Thus, we can definitely judge whether it is a surface localized state.
Contrastingly, if its wave function is algebraically (i.e., weakly) localized
near the step edge, we must be careful in judging it
because a weakly localized state may
merge into bulk states with increasing $L$.
In the case of a weakly localized state, the probability density
near the step edge monotonically decreases with increasing $L$.
If it converges to a finite value in the large-$L$ limit,
the state should be identified as a surface state.
However, the state should be a bulk state
if it becomes vanishingly small in the large-$L$ limit.
This criterion is rephrased as follows:
if the wave function of a weakly localized state is obtained at large but
finite $L$, the state can be identified as a surface state only when
the wave function is normalizable
even after the extrapolation to the limit of $L \to \infty$.

We need to deduce the normalizability of a wave function in the large-$L$ limit
on the basis of numerical data at large but finite $L$.
To do so, the wave function itself, or its probability distribution,
is not convenient as its variation is anisotropic in space.
We thus introduce a partially integrated probability distribution $D(i)$
($i = 1, 2, \dots$) for the two-component wave function
$\mib{\psi}(m,n)= [\psi_{\uparrow}(m,n),\psi_{\downarrow}(m,n)]$
of an eigenstate assuming that it is algebraically localized
near the lower left step edge including the corner site at $(L,1)$.
Here, $D(i)$ is defined as the summation of the probability density
over sites at which the distance from the corner site at $(L,1)$
is smaller than $wi$ but greater than or equal to $w(i-1)$,
where $w$ is an arbitrary length larger than $a$.
That is,
\begin{align}
  D(i) = \sum_{\substack{m,n=1 \\ [wi > R_{m,n} \ge w(i-1)]}}^{L}
         \left(|\psi_{\uparrow}(m,n)|^{2}+|\psi_{\downarrow}(m,n)|^{2}\right) ,
\end{align}
where $R_{m,n}$ denotes the distance between the sites at $(m,n)$ and $(L,1)$.
In determining $R_{m,n}$, the corner site at $(0,1-N)$ should be
identified as that at $(L,1)$ under the shifted periodic boundary condition.

From $D(i)$ calculated for a probable chiral surface state
at large but finite $L$, we deduce its fate in the large-$L$ limit.
If the decrease in $D(i)$ with increasing $i$ is faster than $i^{-1}$,
the state remains normalizable under the extrapolation to the large-$L$ limit,
indicating that it should be identified as a true surface state.
Contrastingly, if the decrease in $D(i)$ is slower than $i^{-1}$,
the state is no longer normalizable under the extrapolation
to the large-$L$ limit, indicating that it merges into bulk states.

Hereafter, we concentrate on low-energy states with negative energy
(i.e., those in the valence band) as they are in one-to-one correspondence
with those with positive energy, and set $k_{x} = 0$ at which
the energy of low-energy states becomes closest to zero energy.
One may think that the fate of a probable chiral surface state can be judged
by observing its energy at $k_{x} = 0$
as it should vanish in the large-$L$ limit.
However, this is prevented by a finite-size gap due to the coupling of
chiral surface states weakly localized near the two different step edges.
As low-energy states are not degenerate, we simply number them
in the valence band from top to bottom
such that the highest and next-highest eigenstates
are respectively called the first and second states.
The first state corresponds to the lowest-energy state closest to zero energy.
Preliminary numerical results suggest that an unpaired chiral surface state
appears whenever $k_{0}a$ exceeds $l \pi/N$ for each positive integer $l$.
Clearly, this condition is different from that for chiral states
along the corresponding screw dislocation.
We thus adopt the working hypothesis that the $j$th state becomes
a chiral surface state when $k_{0}a > j \pi/N$.
For later convenience, we define $\tilde{k}_{0}$ as
\begin{align}
  \tilde{k}_{0} \equiv \frac{k_{0}a}{\pi} ,
\end{align}
in terms of which the condition of $k_{0}a > j \pi/N$ is rewritten as
$\tilde{k}_{0} > j/N$.

\section{Numerical Analysis}

We present the numerical results of $D(i)$ for low-energy states
to examine the working hypothesis for the appearance of
chiral surface states on a step edge.
The states with odd $j$ and those with even $j$ are separately analyzed
because finite-size effects are weaker in the former than in the latter.
We treat the system of $L = 1400$
with step heights of $N = 1$, $2$, $3$, $4$, and $10$.
The parameters are set equal to $B/A = t/A = 0.5$ and $w/a = 4$
with $k_{x} = 0$.

We first consider the case of $N = 1$, in which the hypothesis suggests
that no chiral surface state appears.
Indeed, although the hypothesis superficially indicates that a chiral surface
state appears if $\tilde{k}_{0} > 1$, this is never satisfied
as $\tilde{k}_{0}$ is restricted to be smaller than one.
To examine the absence of a chiral surface state,
we analyze the behavior of $D(i)$ for the first state
at $\tilde{k}_{0} = 0.85$, $0.90$, $0.93$, and $0.96$.
The numerical results are shown in Fig.~3(a),
where solid lines designate the $i^{-1}$ dependence.
We observe that the decrease in $D(i)$ becomes faster with increasing
$\tilde{k}_{0}$ and that its $i$ dependence asymptotically approaches
$i^{-1}$ with increasing $\tilde{k}_{0}$.
That is, the decrease in $D(i)$ does not become faster than $i^{-1}$.
This implies that a chiral surface state does not appear
in the case of $N= 1$,~\cite{comment}
which is consistent with the hypothesis.

We next consider the cases of $N = 2$ and $3$.
In these cases, the hypothesis suggests that the first state,
which should be a bulk state when $\tilde{k}_{0} < 1/N$, becomes
a chiral surface state when $\tilde{k}_{0} > 1/N$.
To examine this, we calculate $D(i)$ for this state
at $\tilde{k}_{0} = 1/N + \delta\tilde{k}_{0}$
with deviations of $\delta\tilde{k}_{0} = 0$, $\pm 0.02$, and $\pm 0.05$.
Figures~3(b) and 3(c) respectively show $D(i)$ in the cases of $N = 2$ and $3$,
where solid lines designate the $i^{-1}$ dependence.
We observe that in both cases, $D(i)$ at $\delta\tilde{k}_{0} = 0$
approximately satisfies $i^{-1}$ in the region of $i > 10$
and that the decrease in $D(i)$ becomes faster
with increasing $\delta\tilde{k}_{0}$.
That is, the decrease in $D(i)$ becomes faster than $i^{-1}$
when $\delta\tilde{k}_{0} > 0$.
This result supports the hypothesis that
the first state becomes a chiral surface state
under the condition of $\tilde{k}_{0} > 1/N$.
The second state in the case of $N = 3$ is considered later.
\begin{figure}[btp]
\begin{center}
\includegraphics[height=6.0cm]{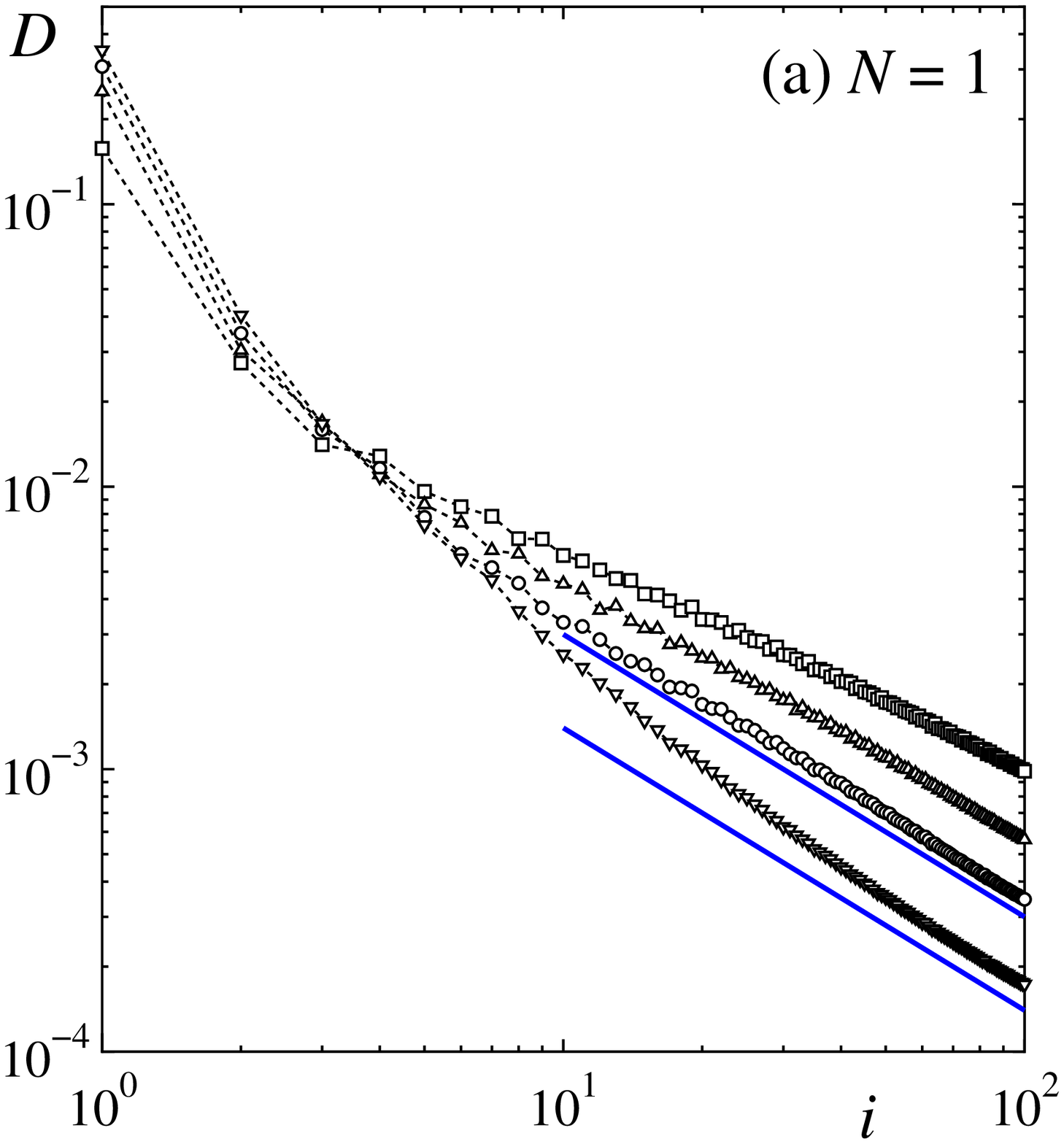}
\includegraphics[height=6.0cm]{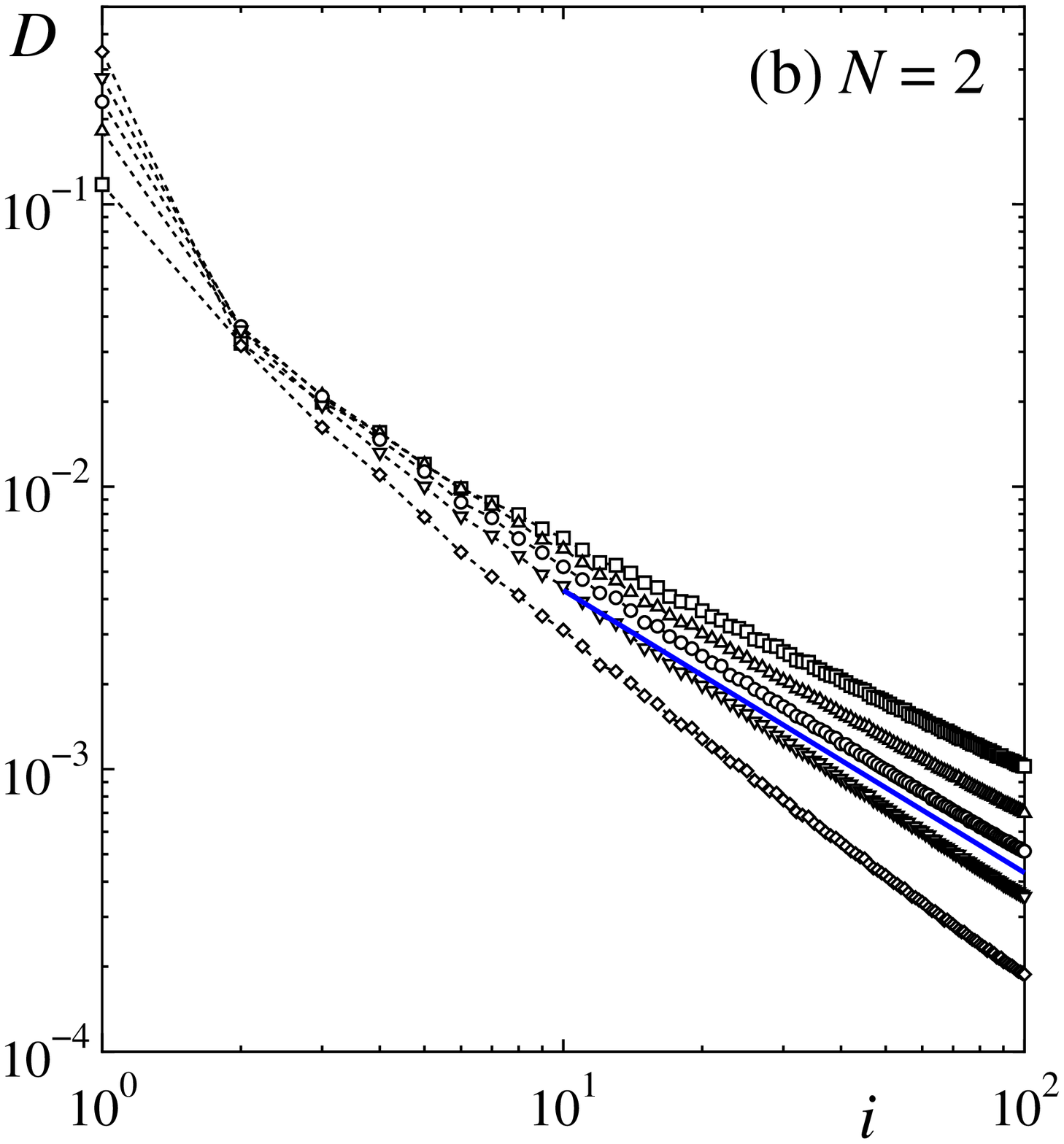}
\includegraphics[height=6.0cm]{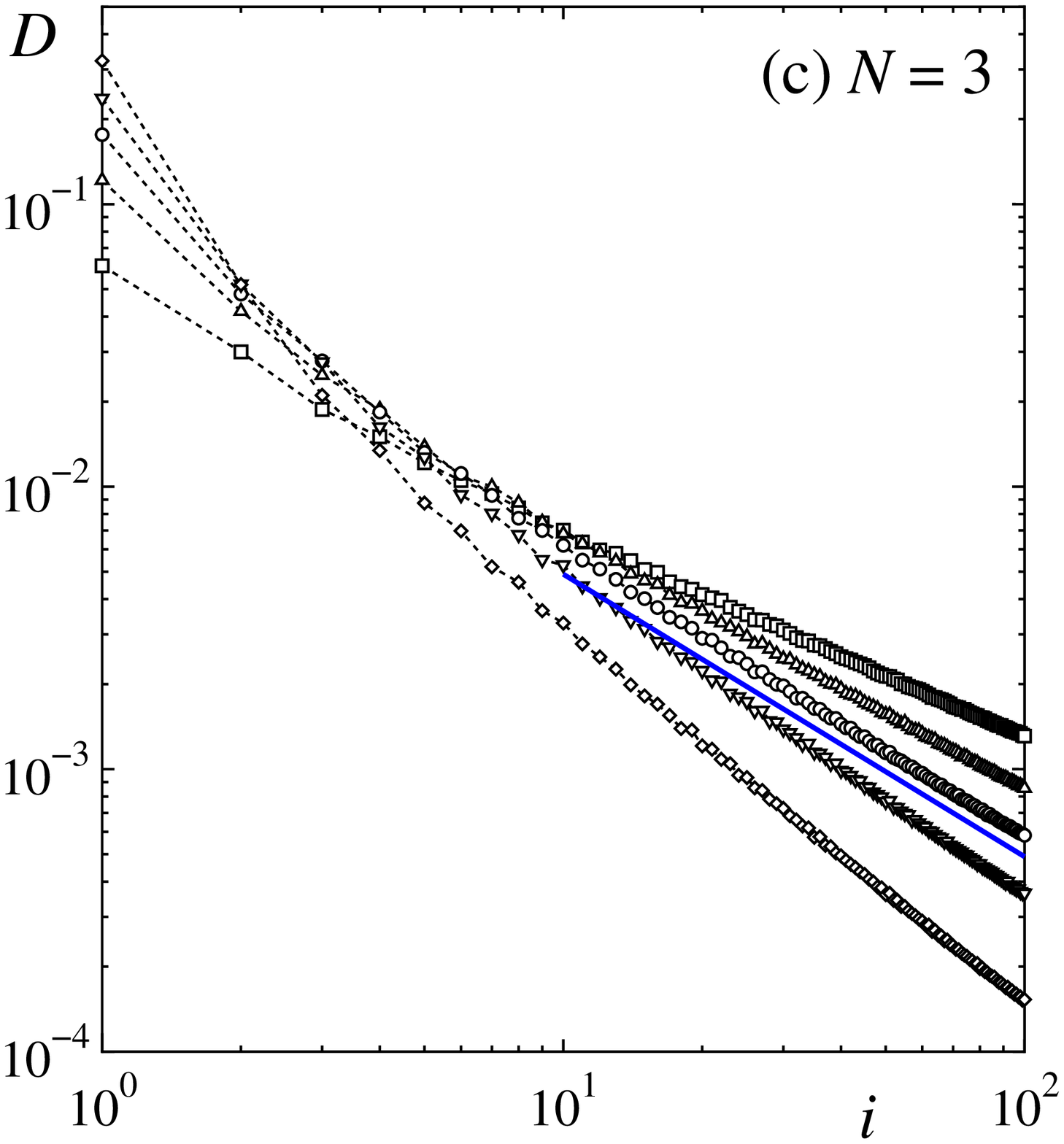}
\end{center}
\caption{(Color online)
$D$ for the first state as functions of $i$ in the cases of (a) $N= 1$,
(b) $N = 2$, and (c) $N = 3$.
In the case of (a) $N= 1$, squares, triangles, circles, and
down-pointing triangles respectively represent $\tilde{k}_{0} = 0.85$, $0.90$,
$0.93$, and $0.96$.
In both the cases of (b) $N = 2$ with $\tilde{k}_{0}=1/2+\delta\tilde{k}_{0}$
and (c) $N = 3$ with $\tilde{k}_{0}=1/3+\delta\tilde{k}_{0}$,
squares, triangles, circles, down-pointing triangles, and diamonds
respectively represent
$\delta\tilde{k}_{0} = -0.05$, $-0.02$, $0$, $0.02$, and $0.05$.
Solid lines designate the $i^{-1}$ dependence.
}
\end{figure}

\begin{figure}[btp]
\begin{center}
\includegraphics[height=6.0cm]{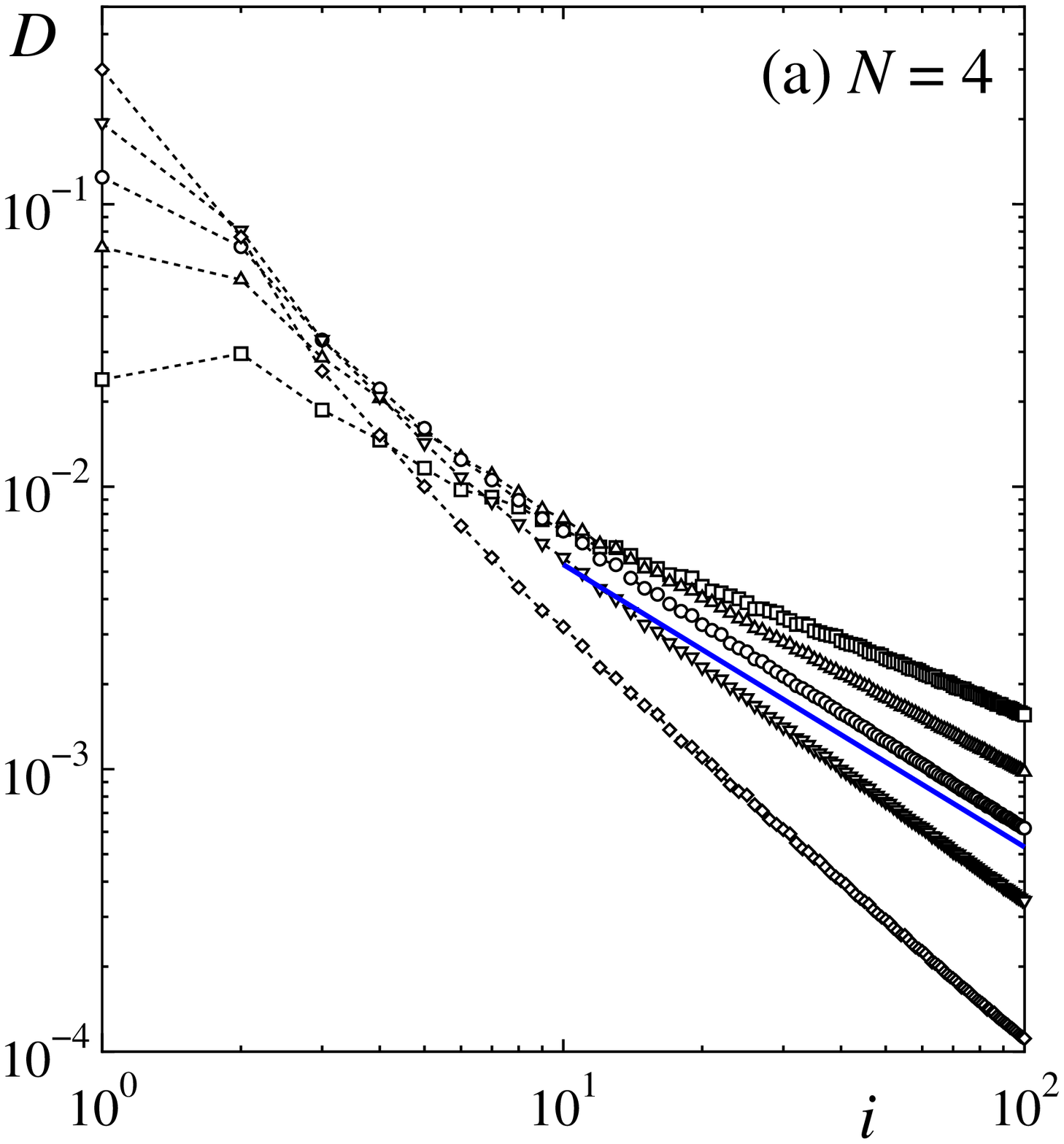}
\includegraphics[height=6.0cm]{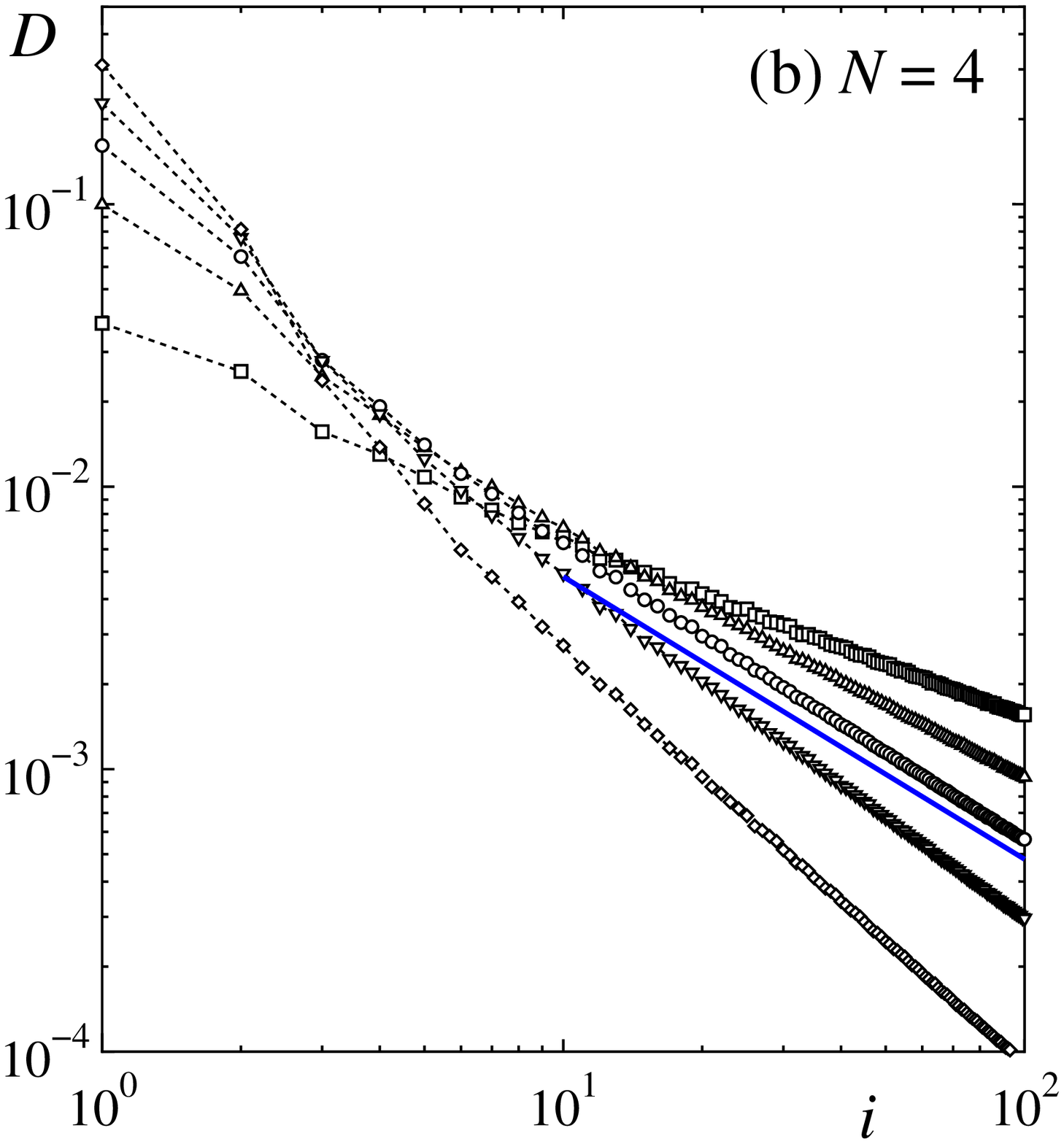}
\end{center}
\caption{(Color online)
$D$ for the (a) first and (b) third states as functions of $i$
in the case of $N = 4$.
In both (a) the first-state case with $\tilde{k}_{0}=1/4+\delta\tilde{k}_{0}$
and (b) the third-state case with $\tilde{k}_{0}=3/4+\delta\tilde{k}_{0}$,
squares, triangles, circles, down-pointing triangles, and diamonds
respectively represent
$\delta\tilde{k}_{0} = -0.05$, $-0.02$, $0$, $0.02$, and $0.05$.
}
\end{figure}
We turn to the case of $N = 4$ with focus on the states with odd $j$,
for which the hypothesis suggests that the first state becomes
a surface state once $\tilde{k}_{0}$ exceeds $1/4$
and then the third state becomes another surface state
when $\tilde{k}_{0}$ exceeds $3/4$.
We thus calculate $D(i)$ for the first state at
$\tilde{k}_{0} = 1/4 + \delta\tilde{k}_{0}$ and that for
the third state at $\tilde{k}_{0} = 3/4 + \delta\tilde{k}_{0}$
with $\delta\tilde{k}_{0} = 0$, $\pm 0.02$, and $\pm 0.05$.
Figures~4(a) and 4(b) respectively show $D(i)$ for the first state
with $\tilde{k}_{0}$ close to $1/4$ and that for the third state
with $\tilde{k}_{0}$ close to $3/4$,
where solid lines designate the $i^{-1}$ dependence.
We observe that in both cases, $D(i)$ at $\delta\tilde{k}_{0} = 0$
approximately satisfies $i^{-1}$ in the region of $i > 10$
and that the decrease in $D(i)$ becomes faster
with increasing $\delta\tilde{k}_{0}$.
That is, the decrease in $D(i)$ becomes faster than $i^{-1}$
when $\delta\tilde{k}_{0} > 0$.
This result again supports the hypothesis.

\begin{figure}[btp]
\begin{center}
\includegraphics[height=6.0cm]{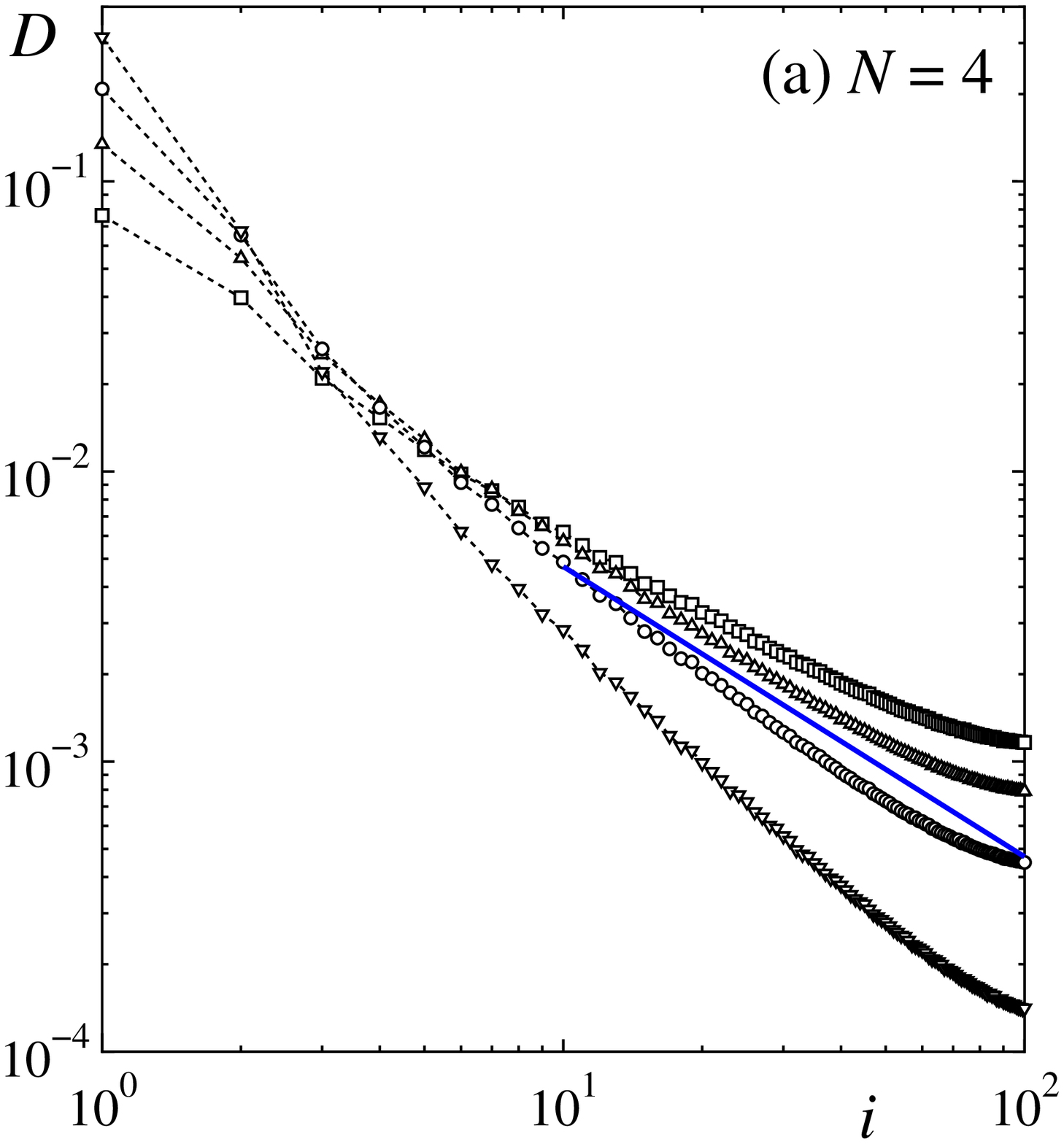}
\includegraphics[height=6.0cm]{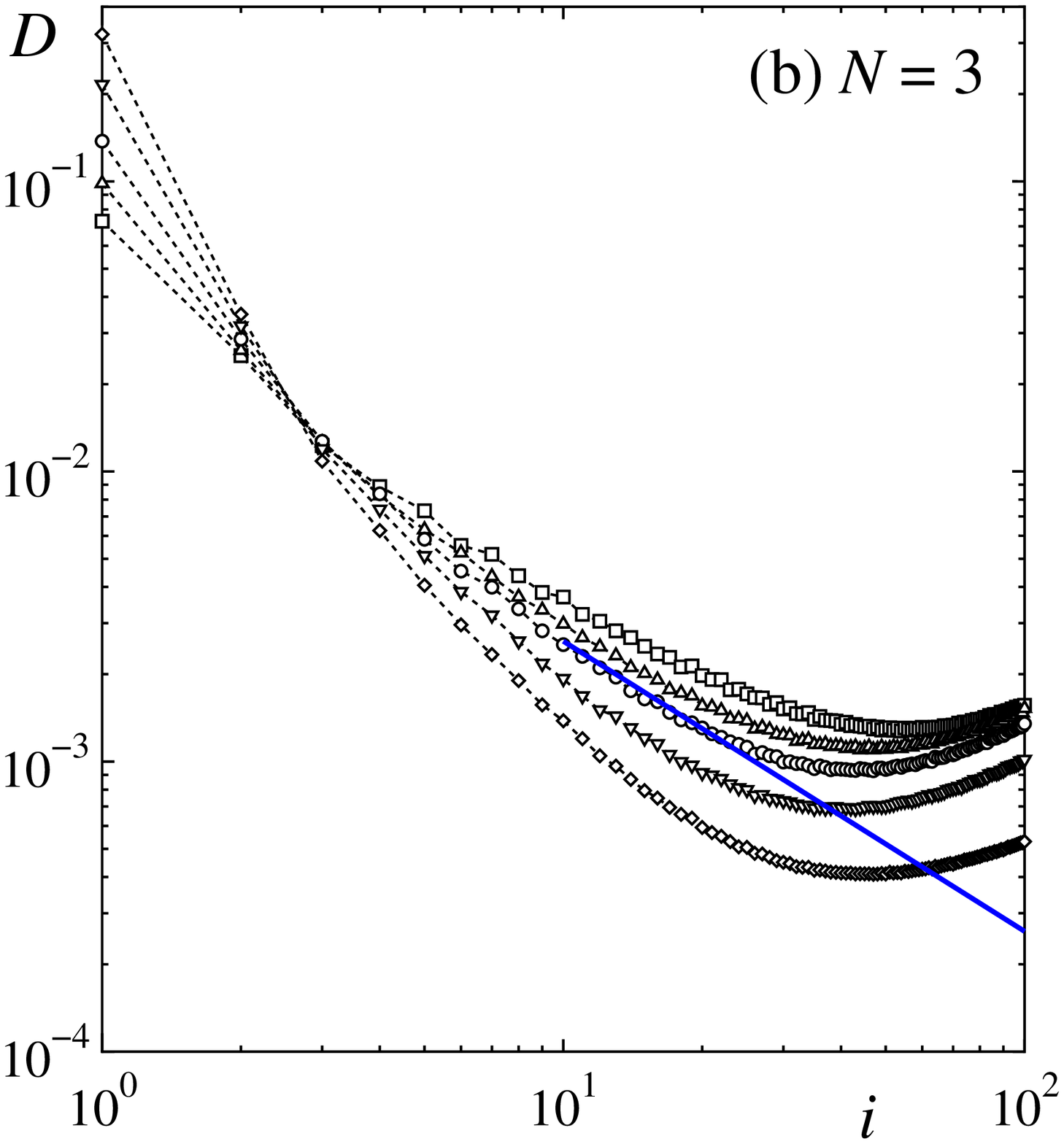}
\end{center}
\caption{(Color online)
$D$ for the second state as functions of $i$ in the cases of (a) $N = 4$
and (b) $N = 3$.
In the case of (a) $N = 4$ with $\tilde{k}_{0} = 2/4 + \delta\tilde{k}_{0}$,
squares, triangles, circles, and down-pointing triangles respectively represent
$\delta\tilde{k}_{0} = -0.02$, $0$, $0.02$, and $0.05$.
In the case of (b) $N = 3$ with $\tilde{k}_{0} = 2/3 + \delta\tilde{k}_{0}$,
squares, triangles, circles, down-pointing triangles, and diamonds
respectively represent
$\delta\tilde{k}_{0} = -0.02$, $0$, $0.02$, $0.05$, and $0.08$.
}
\end{figure}
Let us consider the second state in the cases of $N = 4$ as well as $N = 3$.
Figure~5(a) shows $D(i)$ for this state in the case of $N = 4$
at $\tilde{k}_{0} = 2/4 + \delta\tilde{k}_{0}$ with
$\delta\tilde{k}_{0} = -0.02$, $0$, $0.02$, and $0.05$,
where the solid line designates the $i^{-1}$ dependence.
Remember that the third state becomes a chiral surface state
when $\tilde{k}_{0} > 3/4$,
with the consequence that its energy vanishes in the large-$L$ limit.
This indicates that, at least under the condition of $\tilde{k}_{0} > 3/4$,
the second state should be identified as a chiral surface state
as its energy is smaller than that of the third state.
Hence, the second state must become a chiral surface state at some point
within the interval of $3/4 > \tilde{k}_{0} > 1/4$.
Although $D(i)$ deviates from power-law behavior as $i$ approaches $10^{2}$,
this should be attributed to finite-size effects
in accordance with the argument given above.
From Fig.~5(a), we observe that the decrease in $D(i)$ becomes
faster than $i^{-1}$ when $\delta\tilde{k}_{0} \gtrsim 0$.
This indicates that the second state becomes a chiral surface state
near $\tilde{k}_{0} = 2/4$.
Here, we turn to the second state in the case of $N = 3$.
Figure~5(b) shows $D(i)$ for this state
at $\tilde{k}_{0} = 2/3 + \delta\tilde{k}_{0}$ with
$\delta\tilde{k}_{0} = -0.02$, $0$, $0.02$, $0.05$, and $0.08$,
where the solid line designates the $i^{-1}$ dependence.
As $D(i)$ significantly deviates from power-law behavior
as $i$ approaches $10^{2}$,
it is not easy to obtain a definite conclusion from Fig.~5(b).
However, if this deviation can also be attributed to finite-size effects
as in the case of $N = 4$, we observe that the decrease in $D(i)$
becomes faster than $i^{-1}$ when $\delta\tilde{k}_{0} \gtrsim 0$.
This indicates that the second state becomes a chiral surface state
near $\tilde{k}_{0} = 2/3$.
The results obtained above again support the hypothesis.
As is also shown in Fig.~6(c), finite-size effects are stronger
in the states with even $j$ than in those with odd $j$.
The reason for this is unclear at present.

\begin{figure}[btp]
\begin{center}
\includegraphics[height=6.0cm]{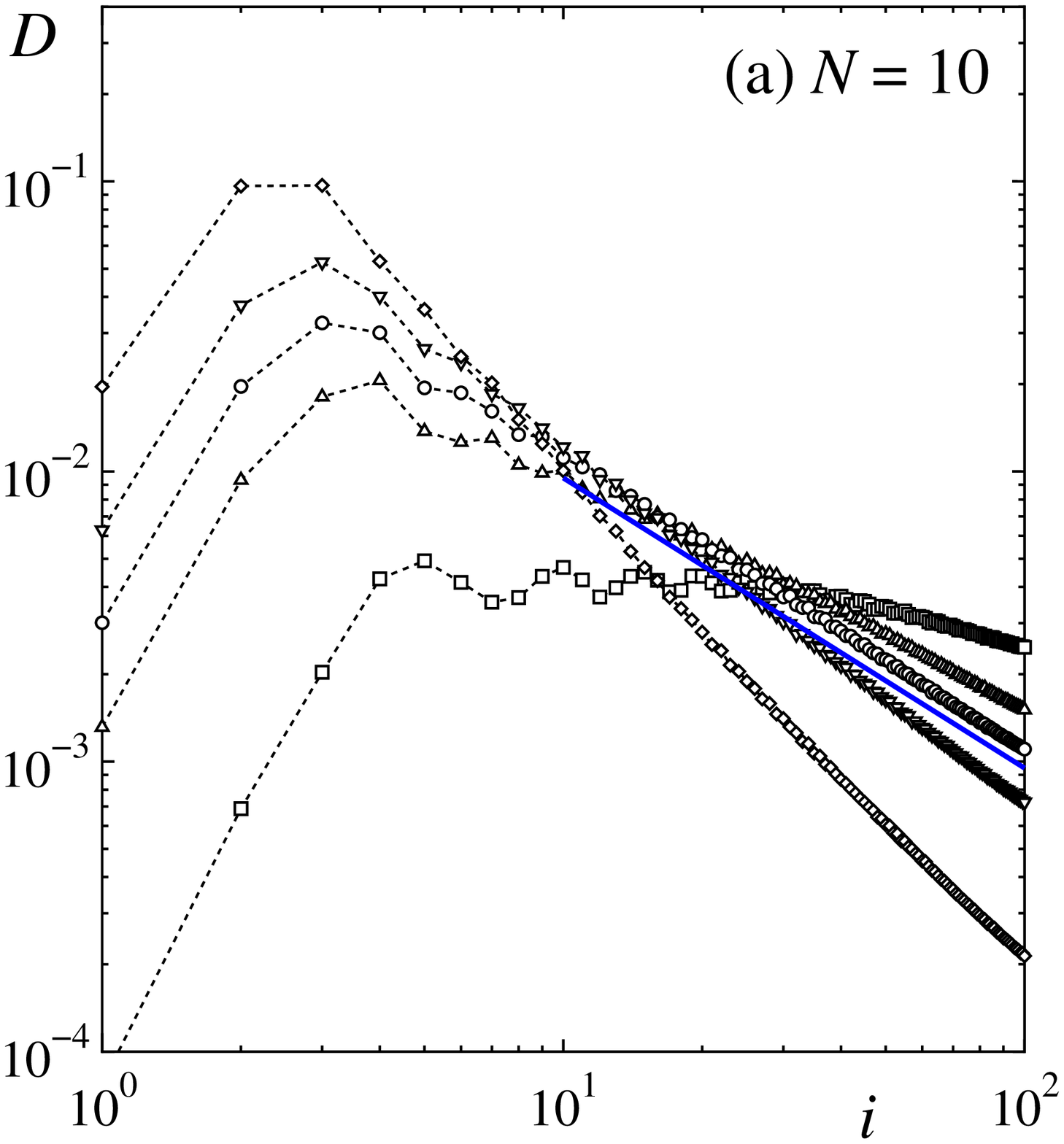}
\includegraphics[height=6.0cm]{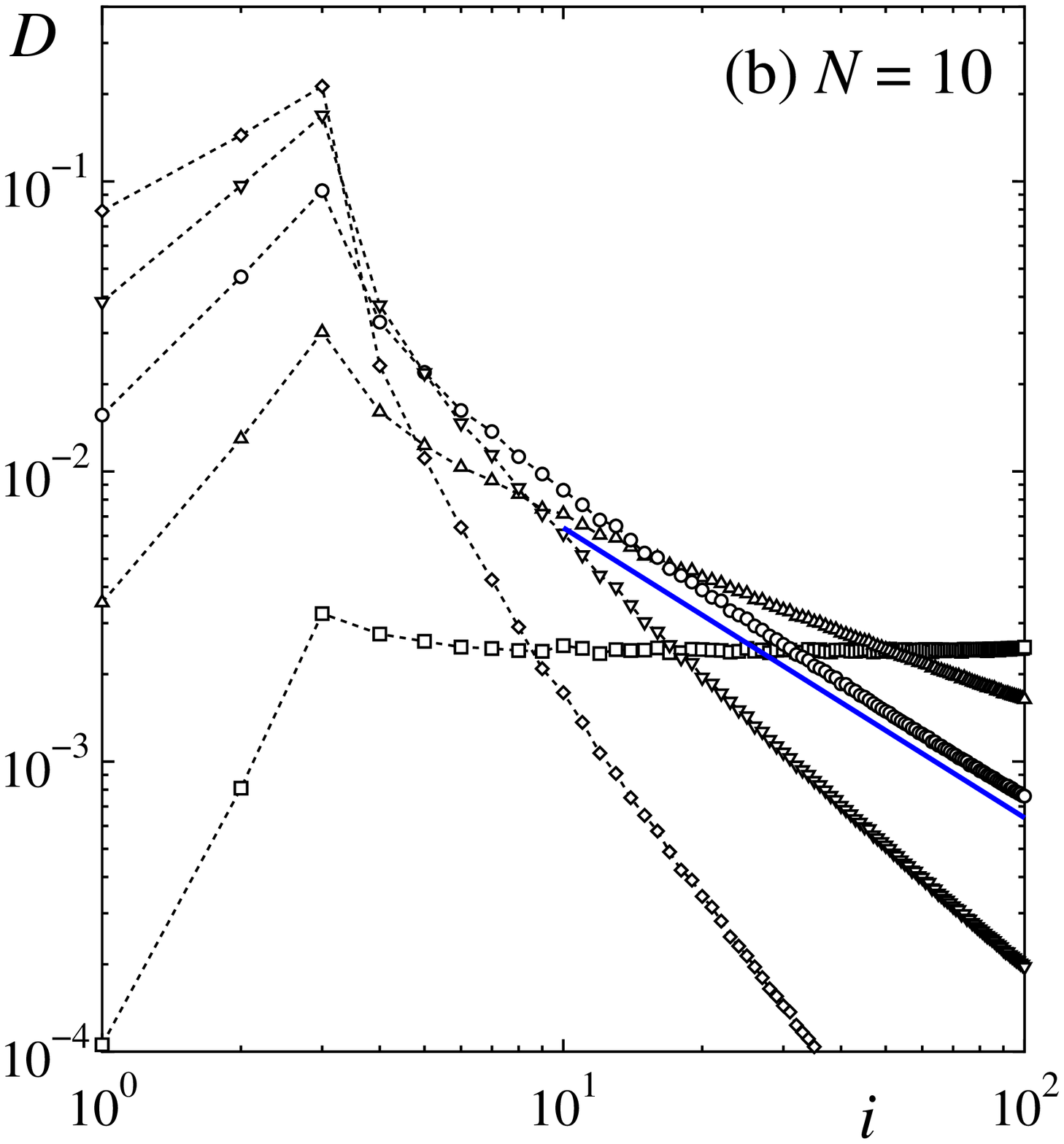}
\includegraphics[height=6.0cm]{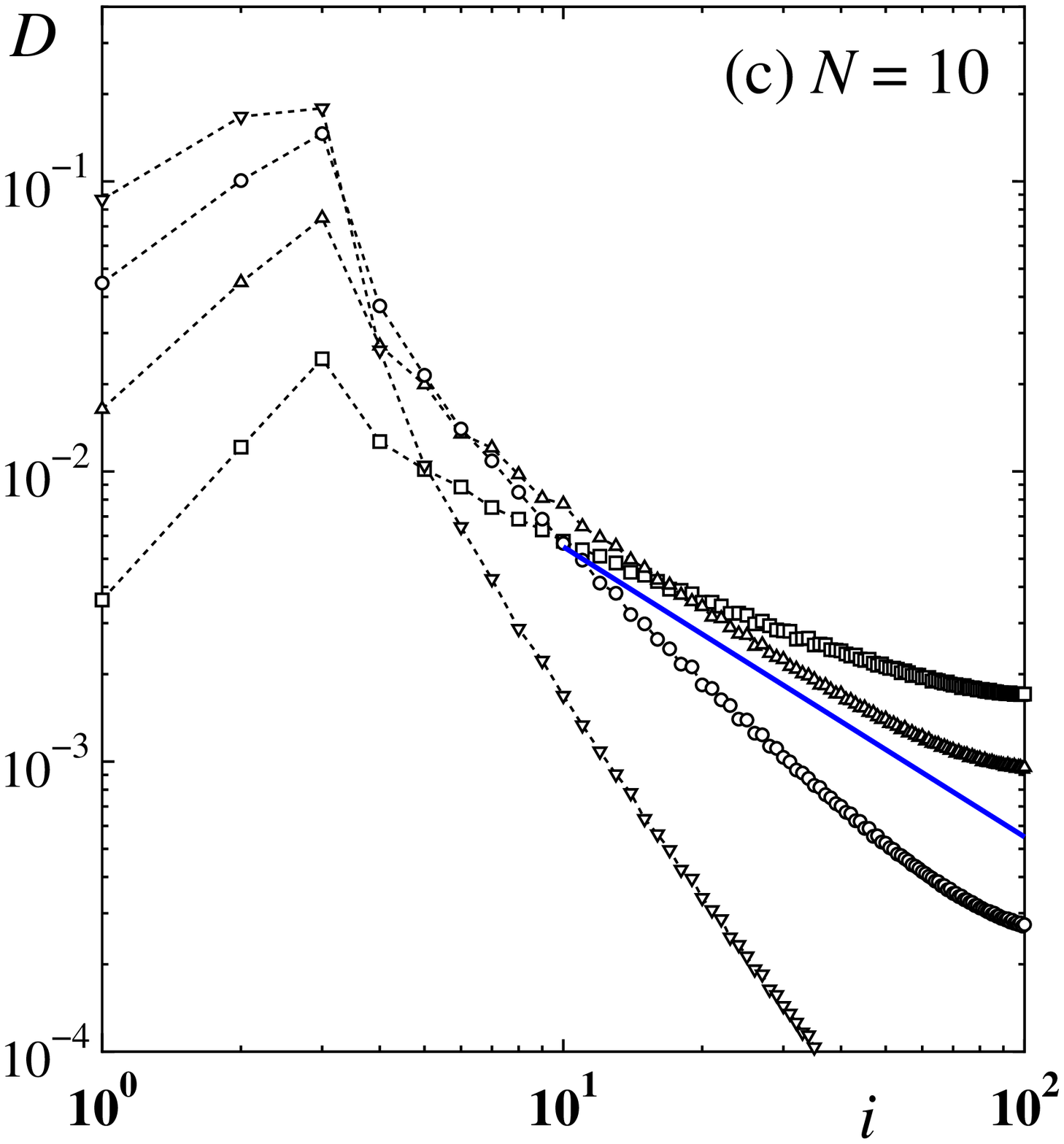}
\end{center}
\caption{(Color online)
$D$ for the (a) first, (b) fifth, and (c) sixth states as functions of $i$
in the case of $N = 10$.
In (a) the first-state case with $\tilde{k}_{0} = 1/10 + \delta\tilde{k}_{0}$,
squares, triangles, circles, down-pointing triangles, and diamonds
respectively represent
$\delta\tilde{k}_{0} = -0.05$, $-0.02$, $-0.01$, $0$, and $0.02$.
In (b) the fifth-state case with $\tilde{k}_{0} = 5/10 + \delta\tilde{k}_{0}$,
squares, triangles, circles, down-pointing triangles, and diamonds
respectively represent
$\delta\tilde{k}_{0} = -0.05$, $-0.02$, $0$, $0.02$, and $0.05$.
In (c) the sixth-state case with $\tilde{k}_{0} = 6/10 + \delta\tilde{k}_{0}$,
squares, triangles, circles, and down-pointing triangles respectively represent
$\delta\tilde{k}_{0} = -0.02$, $0$, $0.02$, and $0.05$.
}
\end{figure}
Finally, we consider the case of $N = 10$
with focus on the states with odd $j$,
for which the hypothesis suggests that the first, third, fifth, seventh,
and ninth states respectively become a chiral surface state
when $\tilde{k}_{0}$ exceeds $1/10$, $3/10$, $5/10$, $7/10$, and $9/10$.
To examine this, we analyze the behavior of $D(i)$ for the first state
with $\tilde{k}_{0}$ close to $1/10$ and
that for the fifth state with $\tilde{k}_{0}$ close to $5/10$.
Figure~6(a) shows $D(i)$ for the first state
at $\tilde{k}_{0} = 1/10 +  \delta\tilde{k}_{0}$ with
$\delta\tilde{k}_{0} = -0.05$, $-0.02$, $-0.01$, $0$, and $0.02$,
where the solid line designates the $i^{-1}$ dependence.
We observe that $D(i)$ approximately satisfies $i^{-1}$ at
$\delta\tilde{k}_{0} = -0.01$
and decreases faster than $i^{-1}$ at $\delta\tilde{k}_{0} = 0$.
This implies that the first state becomes a chiral surface state even when
$\tilde{k}_{0}$ is close to but smaller than $1/10$,
slightly inconsistent with the hypothesis.
However, as the quantitative deviation from the hypothesis is small,
we can say that the hypothesis approximately holds in this case.
Although not shown here, behavior similar to this is also seen in
$D(i)$ for the ninth state with $\tilde{k}_{0}$ close to $9/10$.
The slight inconsistency observed in the first and ninth states
may suggest that the hypothesis holds in a precise manner
only when $N$ is sufficiently small.
Figure~6(b) shows $D(i)$ for the fifth state
at $\tilde{k}_{0} = 5/10 +  \delta\tilde{k}_{0}$ with
$\delta\tilde{k}_{0} = -0.05$, $-0.02$, $0$, $0.02$, and $0.05$,
where the solid line designates the $i^{-1}$ dependence.
The decrease in $D(i)$ becomes faster than $i^{-1}$
when $\delta\tilde{k}_{0} > 0$.
This result supports the hypothesis.
Although not shown here, behavior similar to this is also seen in
$D(i)$ for the third state with $\tilde{k}_{0}$ close to $3/10$
and for the seventh state with $\tilde{k}_{0}$ close to $7/10$.

Here, let us consider the states with even $j$ in the case of $N = 10$.
Figure~6(c) shows $D(i)$ for the sixth state
at $\tilde{k}_{0} = 6/10 +  \delta\tilde{k}_{0}$ with
$\delta\tilde{k}_{0} = -0.02$, $0$, $0.02$, and $0.05$,
where the solid line designates the $i^{-1}$ dependence.
Although $D(i)$ deviates from power-law behavior
as $i$ approaches $10^{2}$ owing to finite-size effects,
we observe that the decrease in $D(i)$ becomes faster than $i^{-1}$
when $\delta\tilde{k}_{0} \gtrsim 0$.
This indicates that the sixth state becomes a chiral surface state
near $\tilde{k}_{0} = 6/10$.
Numerical results for the second, fourth, and eighth states (not shown here)
also indicate that they respectively become a chiral surface state
near $\tilde{k}_{0} = 2/10$, $4/10$, and $8/10$.
These results again support the hypothesis.

\section{Summary}

We numerically studied the behavior of chiral surface states
on a straight step edge with $N$ unit atomic layers
arranged on the flat surface of a Weyl semimetal
possessing a pair of Weyl nodes at $\mib{k}_{\pm} = (0,0,\pm k_{0})$.
We showed that they are algebraically localized near the step edge.
We also showed that the appearance of chiral surface states is approximately
determined by a simple condition characterized by $N$ and $k_{0}$:
an unpaired chiral surface state appears for each positive integer $j$
when $k_{0}a$ exceeds $j\pi/N$, where $a$ is the lattice constant.
Remember that this condition describes the appearance
in the limit of the system size $L$ going to infinity.
If $L$ is large but finite, a weakly localized surface state may appear
even though the condition is not satisfied.
Without satisfying it, however, the probability density
near the step edge monotonically decreases to zero with increasing $L$.
In other words, the condition ensures
the persistence of a chiral surface state in the large-$L$ limit.

Our analysis of chiral surface states is restricted to the case
where the transverse wave number $k_{x}$ is zero.
How do they change if $k_{x}$ deviates from zero?
A chiral surface state at $k_{x} = 0$ likely merges into bulk states
with increasing or decreasing $k_{x}$ in a gradual manner.
If the condition of $k_{0}a > j\pi/N$ is critically satisfied
for a given $j$, we expect that the corresponding chiral surface state
at $k_{x} = 0$ will be fragile against an increase or decrease in $k_{x}$.
That is, it will rapidly merge into bulk states
with increasing or decreasing $k_{x}$.
In contrast, if the condition is sufficiently satisfied
so that $k_{0}a \gg j\pi/N$, the chiral surface state at $k_{x} = 0$
will be robust against an increase or decrease in $k_{x}$.

\section*{Acknowledgment}

This work was supported by JSPS KAKENHI Grant Number 15K05130.


\begin{thebibliography}{99}
\bibitem{comment} An accurate numerical solution is not obtained
within the interval of $0.995 > \tilde{k}_{0} > 0.96$,
possibly owing to the interference between the almost degenerate
first states in the valence and conduction bands.
Hence, this conclusion is tentative in a strict sense.
\end{thebibliography}
\end{document}